
\magnification=1200
\input define2
\v
\centerline{\bf CO Observations of Edge-on Galaxies, V. NGC 5907:}
\centerline{\bf Central Deficiency of Gas in an Sc Galaxy - Merger in the
Bulge?}
\v\v
\cen{\Sofue}
\cen{\iaut}

\v

\cen{(PASJ: Received 1993~~~~~~~~~~~~~~~)}

\v
\v
\cen{\bf Abstract}

The edge-on Sc galaxy NGC 5907 has been observed in the  \co-line emission
using the Nobeyama 45-m telescope.
The radial density distribution between 2 and 13 kpc is well represented by a
superposition of an exponential-law disk of scale radius 3.5 kpc and a ring of
7 kpc radius.
However, no concentration of gas has been observed in the central 2 kpc.
The radial distribution and position-velocity diagram are compared in detail
with those obtained for HI.
We found that the molecular gas dominates in the central 5 kpc region, while HI
does in the outer region, clearly separated by a ``molecular front'' at 5 kpc
radius.
This molecular front will be the place where the phase change from HI to \htwo,
or vice versa, is taking place in a galactic scale.
The central deficiency of the molecular gas is exceptional for a late type
galaxy, which was found for the first time among Sc galaxies, and the second
case after the Sb galaxy M31.
We argue for the possibility of destruction of the nuclear gas disk by a merger
of another galaxy,  and discuss its relation to the outer warping disk.

\v

\sect{\bf I. Introduction}
\v

NGC 5907 is a nearby Sc galaxy with an almost edge-on orientation at an
inclination angle of 88\deg  (Sasaki 1987; Barnaby et al 1992).
Radio continuum maps revealed a thin disk of non-thermal as well as thermal
emissions (e.g., Hummel et al. 1984).
Observations of the HI line emission have shown a large disk of interstellar
gas, which is warping in the outermost regions (Sancisi 1976; Rots 1980; Bosma
1981).
CO ($J=1-0, 2-1$) line emissions have been detected (Braine et al. 1993), while
no mapping observations with a sufficient resolution has been obtained.

This paper is the fifth of a series describing results of a high-resolution
survey of edge-on galaxies in the \co\ line emission using the Nobeyama 45-m
telescope, and we describe the results on NGC 5907.
In our Papers I to IV we presented the results for other edge-on galaxies
(Sofue et al. 1987, 1989, 1990; Sofue and Nakai 1992)

\sect{\bf 2. Observations}
\v

Observations of the \co\ line emission of NGC 5907 were performed in 1992
December using the 45-m telescope of the Nobeyama Radio Observatory.
The parameters for the galaxy are shown in table 1.
The observational parameters are described in Papers III and IV in detail, and
are  summarized in table 2.

\cen{Tables 1, 2}

We used a coordinate system $(X,Y)$, which are defined as the distances along
the major and minor axes from the center position, respectively ($X$ is
positive toward the south-east, and $Y$ is positive toward the north-east).
Observations were made at every $15''$ grid interval along the major axis from
$X=-4'$ to $+4'$, which yielded an effective angular resolution of
$\theta=({\rm HPBW}^2+\Delta X^2)^{1/2}=21''$, except for the edge-most region
at $\vert X \vert > 3'$, where the interval was 30$''$.
Additional $Y$ scans were made at several $X$ positions in order to  confirm
that the CO intensity had a sharp maximum near the adopted major axis.

\sect{\bf 3. Results and Discussion}

\sect{\it 3.1. Spectra and Position-Velocity Diagram}\v

The obtained spectra are shown in figure 1.
The CO emission has been detected at almost all positions except for the
edge-most few points.
Spectra obtained by scans in the $Y$ direction at some $X$ positions indicate
that the emission has a maximum at the adopted galactic plane.
However, the present data are too crude to argue for a thick disk or a halo for
the insufficient coverage of $Y$ distance ($\pm 15''$).
A warping of the outermost part at $\vert X \vert > 4'$ has been reported
(Sasaki 1987), but our data do not cover the area.

\cen{ Fig. 1} 

Using the data we obtained a position-velocity (PV) diagram, which is shown in
figure 2.
The data have been smoothed to a resolution of $20'' \times 20$ \kms.
This yielded an rms noise of about 13 mK in \tmb\ on the PV map, and the
contours are drawn at every 20 mK \tmb.
A rigid rotation feature is observed at $\vert X \vert < 1'$ (3.4 kpc), which
can be attributed to a molecular gas ring of radius 3 to  4 kpc.
Beyond $X\sim \pm 1'$, the rotation velocity is almost constant, indicating a
flat rotation.
By simply tracing the intensity maxima on the PV diagram, we found that a flat
rotation occurs at \vlsr = 460 \kms and 910 \kms, which yields a rotation
velocity of $225(\pm10)$ \kms.

\cen{Fig. 2} 

Not alike as other ``normal'' edge on Sb and Sc galaxies such as the Milky Way
(Dame et al 1990), NGC 891 and NGC 4565 (Papers III and IV), no significant
emission associated with nuclear rotating ring has been detected in the PV
diagram. Such a depression of CO emission near the center is exceptional for an
Sc galaxy.

\sect{\it 3.2. Intensity Distribution and Molecular Mass}\v

The distribution of the integrated  intensity  $I_{\rm CO} = \int T_{\rm mb}dV$
was obtained by integrating the PV diagram in \vlsr,  and is shown in  figure 3
as a function of $X$.
The figure indicates a general concentration of CO gas on a ring at radius
$1-2'$ (3-5 kpc).
The distribution of radio continuum emission at 20 cm along the major axis is
reproduced from Hummel et al. (1984), and is indicated by a dashed line in
figure 4.
We find a global correlation between the CO and continuum.
It is also interesting to note that the galaxy has no radio continuum core,
which is significantly different from other edge-on galaxies like NGC 891 and
NGC 4565 (Hummel et al 1984).

\cen{Fig. 3} 

By integrating the CO emission in figure 2, we estimated the total CO
luminosity of the observed region at $-4' < X < 4'$ and $-7.''5<Y<7.''5$
(within $\pm 13.5$ kpc radius and $\pm 422$ pc thickness) to be  $\Lco = 2.41
(\pm 0.09)\times10^8$ K \kms pc$^2$.
If we assume an H$_2$ column density-to-CO intensity conversion factor of $
C=3.6\times10^{20}$ \htwo\ cm$^{-2}$/K \kms (Sanders et al. 1984), the
luminosity can be related to the \htwo mass as $\Mhtwo=5.7 \Lco$.
Then,  the total mass of H$_2$ gas within the disk is estimated to  be
$\Mhtwo=1.37 (\pm 0.05)\times10^9\Msun$.

\sect{\it 3.3. Radial Distribution and the Molecular Front}\v

By applying the simple decoding method of the integrated CO intensity around
the terminal velocity, as proposed in Papers III and IV,
 we derived the radial distributions of the gaseous surface density as well as
the beam-diluted spatial density.
Here, we took a rotation velocity of $V_{\rm rot}=225$ \kms.
The minimum and maximum velocities of the integration were taken to be  $V_{\rm
min}=190$  and $V_{\rm max}=250$, respectively.
Figure 4 shows thus obtained distribution of the spatial  density of the
molecular gas as a function of the radius.
Here, the ``beam-diluted'' spatial density is defined by the surface density
divided by the linear beam size of the observations.
The density distribution can be approximately represented by a superposition of
an exponential-law disk with the scale radius 3.5 kpc, as indicated by the
dotted line in figure 4, and a ring of radius 7 kpc.
A striking fact in this figure is the sharp depression of the molecular gas
near the center, which will be discussed in section 3.5.

\cen{Fig. 4}

We apply this decoding method to the HI  position-velocity diagram as presented
by Casertano (1983).
In figure 4 we plot thus obtained HI density by a dashed line.
The HI distribution has a broad  ring of 10 kpc radius associated with  an
outskirts reaching as far as 22 kpc radius.
On the other hand,  HI is deficient in the central several kpc.
As shown in figure 4, the HI and \htwo\ gases seem to be present at different
radii, clearly avoiding each other.
Figure 5 plots the mass ratio of \htwo\ gas density to the total (HI+\htwo) gas
density.
The HI gas is dominant in the outer region beyond 5 kpc, while \htwo\ is
dominant in the inner region.
Particularly in the inner 4 kpc region, the  gas  is almost totally molecular.
The ``exchange'' from HI to \htwo\ appears to occur in coincidence with the
molecular gas ring.
As it was discussed in paper IV in detail, we may be able to interpret this
diagram in terms of the existence of a ``molecular front'', at which the HI gas
is converted to \htwo, or vice versa.
This front may be the place where the galactic-scale phase change between HI
and \htwo\ is taking place, and will be deeply coupled with the evolution of
interstellar gas.

\cen{Fig. 5} 

\sect{3.4. CO vs HI}

In figure 6 we superpose the CO PV diagram on the HI PV diagram for the
south-eastern side of NGC 5907, where the HI data are taken from Casertano
1983.
This diagram demonstrates the clear displacement of molecular and atomic
hydrogen gases:
The HI gas composes a large-diameter ring of 12 kpc radius, while molecular gas
is distributed in the inner region at 2 to 10 kpc radius region.
It is remarkable that both gases are distributed clearly avoiding each other:
the inner region is dominated by molecular gas, and the outer region by HI.
In both cases, the rigid-rotation features are due to the molecular and HI gas
rings at these radii.
As was already discussed in section 3.3, it is also clearly observed in this
figure that both emissions have a central depression.

\cen{Fig. 6}

Figure 6 indicates that the CO rotation velocity already attains the maximum at
$r\sim 3-6$ kpc, and is followed by a flat rotation of the outer HI gas.
The fact that the HI and CO rotations attain the same maximum velocity at
$r\sim 5-10$ kpc can be used to argue for the coincidence of the total line
profiles of both emissions:
Figure 7 shows the ``total line profile" in CO for NGC 5907, which was obtained
by integrating the PV diagram in the $X$ direction, and an HI profile as
reproduced from Stavely-Smith and Davies (1988) is superposed.
The figure indicates that the shapes and line widths of the HI and CO line
profiles coincide remarkably well, although the spatial distributions of both
species are significantly displaced.
The total CO velocity width at 20\% level of peak intensity is measured to be
$490 \pm 10$ \kms, and is about equal to that obtained for the HI emission of
480 \kms.
This is consistent with the argument that the total CO line profiles of
galaxies can be used as an alternative to the HI Tully-Fisher (1977) relation,
as has been discussed in the case of NGC 891 (Sofue and Nakai 1992) and for
other galaxies (Sofue 1992; Sch\"oniger and Sofue 1993).

\cen{Fig. 7}

\sect{\it 3.5. Central Molecular Depression: Merger in the Bulge?}

The conspicuous feature observed in figure 4  is the lack in the CO emission
near the center, indicating the deficiency of the interstellar gas near the
nucleus.
Such a clear depression of the central molecular gas is exceptional for an Sc
galaxy, and appears to be the first case among many late type spiral galaxies
so far observed in CO as well as in HI.
In fact most of Sc galaxies show a substantial concentration of CO emission
toward the nucleus, comprising an exponential-law disk (Young and Scoville
1991; Sofue et al 1988; Sofue 1990).

M31 is only one example that was found to exhibit a similar central deficiency
interstellar gas (Brinks and Shane 1984; Koper et al 1994; Sofue and Yoshida
1993):
This giant Sb galaxy is characterized by its 10-kpc ring of interstellar gas
and the ``early-type'' central bulge which is anomalously deficient of
interstellar gas.
It has been suggested that the central gas depression in M31 would have been
caused by a merger of another galaxy, by which the nuclear disk as well as the
inner gas disk were destroyed through the angular momentum transfer as well as
an accretion of  gas from the merger galaxy (Sofue 1994).
The merger in M31 has been indeed suggested to be the case by the discovery of
the double nuclei with the Hubble Space Telescope (Lauer et al 1993).
Therefore, by similarity to M31, we suggest that the nuclear gas disk of NGC
5907 was destroyed by a merger of another galaxy, probably of a companion.
In fact, the outer disk of NGC 5907 is significantly warped (Sancisi 1976), and
could be the evidence for a tidal disturbance which had occurred prior to the
merger during a close encounter of both galaxies.

\v
The author thanks Dr. N. Nakai and the staff at NRO for their help during the
observations.

\sect{\bf References} \v

\r Barnaby, D., and Thronson, Jr. H. A., 1992, \aj, 103, 41.

\r Bosma, A. H. 1980, \aas, 41, 189.

\r Braine, J., Combes, F., Casoli, F., Durpaz, C., Gerin, M., Klein, U.,
Wielebinski, R., and Brouillet, N. 1993, \aas, 97, 887.

\r Brinks, E. and Shane, W. W.  1984,  \aa, 55, 179.

\r Casertano, S. 1983, \mn, 203, 735.

\r Dame, T. M., Ungerechts, H., Cohen, R. S., de Geus, E. J., Grenier, I. A.,
May, J., Murphy, D. C., Nyman, L. -A, and Thaddeus, P.  1987, \apj, {\bf 322},
706.

\r Hummel, E., Sancisi, R., and Ekers, R. D. 1984, \aa, {\bf 133}, 1.

\r Lauer, T., Faber, S. et al. 1993, \aj, October issue, in press.

\r Koper, E., Dame, T. M., Israel, F. P., Thaddeus, P. 1991, \apjl, 383, L11.

\r Rots, A. 1980, \aas, {\bf 41}, 189.


\r Sancisi, R. 1976, in {\it Topics in Interstellar Matter}, ed. H. van Woerden
(D. Reidel Pub. Co., Dordrecht), p. 255.

\r Sasaki, T. 1987, \pa, 39, 849.

\r Sanders, D. B., Solomon, P. M., and Scoville, N. Z. 1984, \apj, {\bf 276},
182.

\r Sch\"oniger, F., and \so\ 1993, \aa, in press.

\r \so\ 1991, \pa, 43, 671.
\r
\r \so\ 1992, \pal, 44, L231.

\r \so\ 1994, ApJ, March issue, in press.

\r \so, Doi, M.,  Ishizuki, S.,  \na, and \ha\ 1988, \pa, 40, 511.

\r \so, \ha, and \na\   1989, \pa, {\bf 41}, 937 (Paper II).

\r \so, and \na\ 1993, \pa, 45, 139 (Paper III).

\r \so, and \na\ 1993, \pa, in press  (Paper IV).

\r \so, \na, and \ha\  1987, \pa, {\bf 39}, 47 (Paper I).

\r \so, and Yoshida, S. 1993, ApjL, in press

\r Staveley-Smith, L., Davies, R. D. 1988, MNRAS 231, 833.

\r Tully, B., and Fisher, J. R. 1977, \aa, {\bf 54}, 661.

\r Young, J., Scoville, N. Z.  1991, ARAA 29,  581.

\endpage

\settabs 3 \columns
\noindent{Table 1. Parameters for NGC 5907.}
\v

\hrule\v

\+ Galaxy type  \dotfill &  Sc; edge-on \cr

\+ Center position ($X=0'',~Y=0''$)&& (NED$^\dagger$) \cr

\+ ~~~  R.A.$_{1950}$   \dotfill& $15^{\rm h} 14^{\rm m} 34^{\rm s}.80$ \cr

\+ ~~~  Decl.$_{1950}$  \dotfill& $56\Deg 30' 33.''0$ \cr

\+ Major axis P.A.   \dotfill& 156\deg & (Sasaki 1987) \cr

\+ Inclination angle  \dotfill& 88\deg  & (Sasaki 1987) \cr

\+  Systemic \vlsr  \dotfill& 685 ($\pm 5$) \kms & (Present CO result) \cr

\+ \vrot   \dotfill& 225 ($\pm 5$) \kms  & (Present CO result) \cr

\+ Distance 	\dotfill& 11.6 Mpc & (Sch\"oniger and Sofue 1993)  \cr

\+ CO luminosity$^\ddagger$  \dotfill  & $ 2.43 (\pm 0.09)\times10^8$ K \kms
pc$^2$ \cr

\+ \htwo mass \dotfill & $1.37 (\pm 0.05)\times10^9\Msun$ & Conversion factor
from Sanders et al (19874) \cr

\v
\hrule
\v
$\dagger$ NASA/IPAC Extragalactic Database (NED) (1992) (operated by JPL,
Cal. I. Tech. under contract by NASA).

$\ddagger$ Values in a ``thin'' disk of  27 kpc $\times$ 844 pc

\vskip 5mm

\settabs 2 \columns
\noindent{Table 2. Observational Parameters.}
\v

\hrule\v
\v
\+ Telescope \dotfill & Nobeyama 45-m telescope \cr

\+ Angular resolution (HPBW) \dotfill & 15$''$ (844 pc) \cr

\+ Pointing accuracy \dotfill & $\pm 3''$ (calibrated by SiO masers)\cr

\+ Aperture efficiency \dotfill & 0.35 \cr

\+ Main-beam efficiency \dotfill & 0.50 \cr

\+ Integration time per point \dotfill & 5 to 10 min. \cr

\+ Velocity resolution \dotfill & 10 \kms (32-ch binding from 2048-ch AOS) \cr

\+ Receiver system temperature \dotfill & 400 - 500 K (SIS receiver; SSB) \cr

\+ Rms of each spectra (10 \kms resol.) \dotfill & 20 mK in \tmb \cr

\+ Rms on PV diagram ($20''\times 20$ \kms) \dotfill & 12 mK \cr.

\v
\hrule

\endpage

\noindent {\bf Figure Captions} \v

Fig. 1: \co\ line spectra of NGC 5907 as observed with the 45-m telescope at
Nobeyama.

Fig. 2: A Position-velocity $(X-V)$ diagram along the major axis of NGC 5907.
The resolution is $20'' \times 20$ \kms,  and the rms noise of
the map is about 13 mK in \tmb. 
Contours are drawn at every 20 mK in \tmb\ starting at 20 mK, and the peak
intensity in the map is 245 mK \tmb.

Fig. 3: \ico\ distribution along the major axis $X$ (full line).
The resolution in the $X$ direction is $20''$.
Intensity of 21-cm continuum emission (resolution 28$''$: Hummel et al 1984)
is superposed by a dashed line.

Fig. 4: Beam-diluted spatial densities of molecular (thick line) and neutral
hydrogen (dashed line) gases
(in H \cc).
The thin line shows an exponential-law disk with a scale radius 3.5 kpc fitted
to the molecular gas distribution.

Fig. 5: The mass ratio of \htwo\ to total (HI+\htwo) gas density as a
function of the galactocentric distance.

Fig. 6: Comparison of the CO and HI distributions on the PV diagram.
 A displacement of both species is shown here most clearly.

Fig. 7: A total line profile of the \co\ emission along the major axis.
Velocity resolution in this diagram is 10 \kms.
An HI total line profile is shown by the dashed line (Staveley-Smith and Davies
1988).

\bye